\documentclass[aps,pre,preprint,amssymb,amsmath,nofootinbib, superscriptaddress]{revtex4}
\usepackage{dcolumn}
\usepackage{bm}
\usepackage{natbib}
\usepackage{xr}
\usepackage{graphicx}
\usepackage{amsmath}
\usepackage{amsfonts}
\usepackage{amssymb}
\usepackage{amsthm}
\usepackage{bbold}
\usepackage[usenames,dvipsnames]{color}
\usepackage{listings}
\usepackage{rotfloat}
\usepackage{color}
\definecolor{darkblue}{RGB}{0, 0, 128}
\usepackage[colorlinks=true, linkcolor=darkblue, citecolor=darkblue]{hyperref}
\usepackage{algorithm, algorithmic}

\begin{document}

\title{Simulation of stochastic network dynamics via entropic matching}

\author{Tiago Ramalho}
\email{tiago.ramalho@physik.uni-muenchen.de}
\affiliation{Arnold Sommerfeld Center for Theoretical Physics (ASC) and Center for Nanoscience (CeNS), \\ LMU M\"unchen, Theresienstra{\ss}e 37, 80333 M\"unchen, Germany}

\author{Marco Selig}
\affiliation{Max Planck Institute for Astrophysics (MPA), Karl-Schwarzschild-Stra{\ss}e 1, 85741 Garching, Germany}

\author{Ulrich Gerland}
\affiliation{Arnold Sommerfeld Center for Theoretical Physics (ASC) and Center for Nanoscience (CeNS), \\ LMU M\"unchen, Theresienstra{\ss}e 37, 80333 M\"unchen, Germany}

\author{Torsten A. En{\ss}lin}
\affiliation{Max Planck Institute for Astrophysics (MPA), Karl-Schwarzschild-Stra{\ss}e 1, 85741 Garching, Germany}

\begin{abstract}
The simulation of complex stochastic network dynamics arising, for instance, from models of coupled biomolecular processes remains computationally challenging. Often, the necessity to scan a models' dynamics over a large parameter space renders full-fledged stochastic simulations impractical, motivating approximation schemes. Here we propose an approximation scheme which improves upon the standard linear noise approximation while retaining similar computational complexity. The underlying idea is to minimize, at each time step, the Kullback-Leibler divergence between the true time evolved probability distribution and a Gaussian approximation (entropic matching). This condition leads to ordinary differential equations for the mean and the covariance matrix of the Gaussian. For cases of weak nonlinearity, the method is more accurate than the linear method when both are compared to stochastic simulations. 
\end{abstract}
\maketitle

\section{Introduction}

Stochastic dynamical systems arise in many areas of science, with much recent interest focusing on biomolecular reaction networks that control the behavior of individual cells \cite{Raser:2005gr, Eldar:2010kk, Balazsi:2011bw, Munsky:2012ie}. For instance, the expression levels of different genes in a single cell are often coupled by direct or indirect regulation and are subject to intrinsic (reaction) noise as well as extrinsic (parametric) noise \cite{Elowitz:2002wq}. Models of such systems can be based on a wide range of different approaches and mathematical tools \cite{de_jong_modeling_2002, wilkinson_stochastic_2009}. A full treatment of noise effects on the single-molecule level requires an approach based on the chemical Master equation \cite{van_kampen_2007}. However, many noise phenomena can also be understood and quantitatively analyzed within a more coarse-grained continuum description based on stochastic differential equations or, equivalently, the associated Fokker-Planck equations \cite{wilkinson_stochastic_2009, van_kampen_2007, gardiner_stochastic_2010}. For instance, such a description is adequate to describe the excitable system behavior of a natural genetic circuit whose excitations are triggered by noise \cite{Suel:2006ea}, or the synchronizing effect of a coupling between noisy synthetic ``repressilators'' \cite{elowitz_synthetic_2000, GarciaOjalvo:2004vg}. 

A recurrent problem in the analysis of these systems is that the quantitative behavior of the theoretical models must be computed not only once, but for a large number of different parameter values. One common situation is that, for a given  model, one would like to determine a ``phase diagram'' of system behaviors over the entire parameter space, in order to understand the model on a theoretical level. Similarly, when models are leveraged for the interpretation of experiments, model fitting to data is required to infer system parameters or to discriminate between different variants of the model. In quantitative biology, there is currently a clear need for systematic and efficient methods to reverse engineer models from a limited number of noisy experimental observations \cite{Prill:2011cc, Steiert:2012ke}. Any such method requires as an essential ingredient an efficient technique for the forward simulation of the observables. For discrete stochastic models based on the chemical Master equation, recent progress in this direction has been made by using spectral methods \cite{walczak_stochastic_2009} or efficient cutoff schemes \cite{barzel_binomial_2011}. Also, efficient sampling techniques have been developed for rare event problems \cite{allen_sampling_2005, Becker:2012fl}. Here, we focus on continuum models and develop an approximation scheme that can efficiently capture the stochastic dynamics of the ``typical events'' even in larger dynamical systems.   

For a given nonlinear stochastic system, we consider the dynamics of its probability distribution over state space. Our scheme is reminiscent of the linear noise approximation \cite{van_kampen_2007, gardiner_stochastic_2010} in that it approximates the probability distribution at each point in time by a Gaussian distribution. The essential difference is in the criterion used to match the Gaussian approximation to the true distribution. In our method, we take a more global stance at the matching process, by invoking the Kullback-Leibler divergence~\cite{Kullback1951} as accuracy measure. We minimize the Kullback-Leibler divergence at each point in time to obtain a set of ordinary differential equations for the parameters of the Gaussian distribution which best approximates, in this sense, the current probabilistic state of the system. We refer to our matching condition as entropic matching. It was recently proposed as a general method to construct simulation schemes for partial differential equations \cite{2012arXiv1206.4229E}. Here, we develop and test this idea for the case of stochastic differential and Fokker-Planck equations. For the examples considered here, we find the method to be more accurate and robust than the linear noise approximation, while the computational complexity is comparable. In the following, we first formulate the theoretical problem, then describe the method in detail, and finally illustrate and test  it on simple example systems.

\section{Problem statement}
At the most fundamental level a biochemical reaction network can be represented by a stochastic jump process \cite{van_kampen_2007}. A specific system state is described by a vector $n=\{n_1,...,n_N\}$ which represents the number of molecules for each of the $N$ different molecular species. The transition rates are represented by a linear operator $W$, where an element $W_{n', n}$ is the transition rate for a specific jump (from state $n$ to $n'$). The nonlinearity of the underlying dynamical system is reflected in the nonlinear dependence of $W_{n', n}$ on $n$, $n'$. 
At each time step the probability to be in a given system state is represented by $P(n,t)$ and its evolution is given by the chemical Master equation 
\begin{equation}
\label{eq:MASTER}
\partial_t P(n,t) = \sum_{n'} (W_{n', n} P(n,t) - W_{n, n'} P(n', t)) \;.
\end{equation}
Generally it is not possible to solve this equation analytically, but there are a number of approaches to efficiently solve it numerically~\cite{walczak_stochastic_2009, barzel_binomial_2011}. However, even these efficient approaches become prohibitive when the total number of molecules $\sum_i n_i$ is large. 

For a large number of molecules, a well-suited approximation of the discrete stochastic process is by a continuous drift and diffusion process \cite{van_kampen_2007, gardiner_stochastic_2010}. This description bridges the microscopic stochastic picture with the macroscopic deterministic picture and provides important insights for a regime in which particle numbers are high but stochasticity still plays an important role in the system dynamics~\cite{wilkinson_stochastic_2009}. In this framework the system state is described by a set of continuous concentrations $c=\{c_1,...,c_N\}$. The interactions between the molecular species are specified by a set of nonlinear dynamical equations, which depend on numerous parameters denoted henceforth as $a$. 

The dynamical equations can be written as stochastic differential equations (SDE)
\begin{equation}
    \label{eq:SDE}
    \dot{c}(t) = \frac{d}{dt}\,c(t) = f(c(t), a) + \eta(t)
    \;,
\end{equation}
where $f$ is the nonlinear function regulating each node of the biomolecular network and $\eta$ represents a random process characterizing the system's intrinsic fluctuations. We can think of a single trajectory obtained by solving Eq.~\eqref{eq:SDE} (with one realization of the noise process $\eta(t)$) as the biochemical trajectory of a single cell, and an ensemble of such trajectories as the corresponding dynamics for a population of cells. This picture relies on the assumption that different cells' trajectories are uncorrelated, i.e. that there is no intercellular communication. Within the framework of SDE's, extrinsic cell-to-cell noise would need to be accounted for by specifying a probability distribution for the parameters $a$, which can either be static or have a dynamics by itself. For simplicity, we will ignore extrinsic noise in the following. From the SDE system \eqref{eq:SDE} one can then directly obtain \cite{van_kampen_2007, gardiner_stochastic_2010} a Fokker-Plank Equation (FPE) of the form 
\begin{align}
   \label{eq:FPE}
    \partial_t P(c,t) &= -\partial_{c} \left[ f(c, a)\, P(c, t) \right] + \tfrac{1}{2} \partial_{c}^2 \left[ \mathcal{X} \,P(c, t) \right]
    \;,
\end{align}
where $\mathcal{X}$ represents the covariance of the random process denoted by $\eta$ in Eq.~\eqref{eq:SDE}. Note that $\mathcal{X}$ generally depends on $c$. We have suppressed this dependence to simplify the notation. This FPE can be solved analytically only in a very restricted number of cases, such as when $f$ and $\mathcal{X}$ are linear. Note that instead of deriving the FPE from a phenomenological set of SDE's, one could have also started from the chemical Master equation~\eqref{eq:MASTER} and applied an appropriate limiting procedure to recover the FPE \cite{van_kampen_2007}.

The FPE could be solved numerically using traditional algorithms for partial differential equations such as finite elements and finite differences. Since these methods rely on spatial and temporal discretization, the grid size grows exponentially with the number of dimensions which makes them impractical for systems with several chemical species. Another possibility is to simulate many individual trajectories of the SDE system, Eq.~\eqref{eq:SDE}, using numerical algorithms for stochastic differential equations \cite{kloeden_numerical_1992}, and reconstruct the probability density function from these trajectories. However, as dimensionality increases, each individual path will cover an increasingly smaller portion of phase space, which means a huge number of paths must be simulated to obtain sufficient statistics. Therefore, this method is also impractical for larger systems. However, in our test applications below (which are small systems), we use this exact approach as a reference for our approximation scheme.

The method here proposed relies on describing the probability distribution by only a small set of parameters, which evolve through time. To implement this idea, let us approximate the distribution $P(c,t|a)$ at each time point by a Gaussian distribution, 
\begin{align}
    \label{eq:popgauss}
    P(c,t|a) &= \mathcal{G}(c-\bar{c}(t),C(t)),\\
    \label{eq:Gaussian}
    \mathcal{G}(c-\bar{c}, C) &= \frac{1}{\sqrt{|2\pi\,C|}} e^{-\frac{1}{2}\,(c-\bar{c})^\dagger C^{-1}(c-\bar{c})}.
\end{align}
Here, $|C|=\mathrm{det}(C)$ and $c^\dagger d = \sum_{i=1}^N c_i\,d_i$ is a scalar product over the space of network state vectors. In the Gaussian approximation, knowledge of the temporal dynamics of its average ${\bar c}$ and covariance matrix $C$ is sufficient, which enormously simplifies all calculations. The proposed method consists of calculating $\bar{c}(t)$ and $C(t)$ such that the Gaussian approximation best represents $P(c|a, t)$ in an information theoretic sense. The resulting equations for $\bar{c}(t)$ and $C(t)$ can be efficiently solved numerically. How well the true non-Gaussian probability distribution can be approximated by distribution \eqref{eq:popgauss} is of course critical to the performance of our approach, as will become clear in our test applications.

\section{Method}
\subsection{Background}

Since we are approximating the probability distribution by the Gaussian \eqref{eq:popgauss}, only the time evolution of the first two moments needs to be calculated. Hence we seek closed ordinary differential equations (ODEs) for the average and the covariance matrix, instead of solving a partial differential equation for the full probability distribution. These ODEs are specified by two functions, $f_{\bar{c}}$ and $f_C$ such that 
\begin{align}\label{eq:cCevolution}
    \dot{\bar{c}}(t)&=f_{\bar{c}}\left(\bar{c}(t),C(t)\right) \;, \nonumber \\
    \dot{C}(t)&=f_C\left(\bar{c}(t),C(t)\right) \;.
\end{align}
The simplest approach is the so-called linear noise approximation \cite{van_kampen_2007}, where the evolution of the covariance is determined strictly locally, by the Jacobian $J$ of the function $f$ at the current average concentration $\bar{c}(t)$, 
\begin{align}
\label{eq:linear}
\dot{\bar{c}} (t) &= f(\bar{c}) \;, \nonumber \\
\dot{C}(t) &= J(t) C(t) + C(t) J^T(t) + \mathcal{X}(t) \;.
\end{align}
In our test applications below, we use this linear noise approximation as a reference for comparison to our proposed method.

For our method, we now derive evolution equations of the general form \eqref{eq:cCevolution} that go beyond the strictly local treatment \eqref{eq:linear} of the linear noise approximation. Thereby, we seek to better preserve the characteristics of the original nonlinear stochastic system. 

\subsection{Proposed method}

We will construct $f_{\bar{c}}$ and $f_C$ using the concept of entropic matching for evolving distributions \cite{2012arXiv1206.4229E}. Roughly, our procedure is to assume an initial Gaussian distribution for $c$, evolve it according to the nonlinear function $f$, and then find the Gaussian distribution which best matches it in an information theoretic sense. A Gaussian distribution $P(c)=\mathcal{G}(c-\bar{c},\,C)$ characterized by $\bar{c}=\bar{c}(t)$ and $C=C(t)$ will evolve according to the dynamics described by the nonlinear function $f$ and the noise covariance $\mathcal{X}$. 
For now, we focus on the deterministic part of this evolution and add the intrinsic noise below. After an infinitesimal time step $\delta t$, the evolved distribution function $P'(c')$ will be 
\begin{equation}
\label{eq:evdet}
    P'(c')= \mathcal{G}(c-\bar{c},C)|_{c=c'-\delta t \,\dot{c}}\left|\frac{d c}{d c'}\right| 
\end{equation}
according to the rule of change of variables for probabilities, where $c(t+\delta t) = c' = c + \delta t \,\dot{c}$ and thus $|dc'/dc|_{c=c'}=|1+\delta t \; df/dc|_{c=c'}$ to leading order in $\delta t$. $P'(c')$ is a normalized probability distribution, although it is in general not of Gaussian functional form. Obviously a representation such as $P'(c')= \mathcal{G}(c'-\bar{c}',\,C')$ with an updated $\bar{c}'=\bar{c}(t')$ and $C'=C(t')$ at $t'=t+\delta t$ is no longer exact. However, we can determine values for $\bar{c}'$ and $C'$ such that the information content of $P'(c')$ is represented as well as possible. 

Information theoretical considerations \cite{1957PhRv..106..620J, 1957PhRv..108..171J, 2003prth.book.....J} single out the Maximum Entropy principle for this process. This principle is equivalent to requiring a minimal Kullback-Leibler (KL) divergence \cite{Kullback1951} of $\mathcal{G}(c'-\bar{c}',\,C')$ to $P'$, or a minimal relative Gibbs free energy \cite{2010PhRvE..82e1112E}. All these measures (relative entropy, Kullback-Leibler divergence, and Gibbs free energy) can be regarded as a  measure of the information theoretical distance between two distributions, although they are not a metric in the strict mathematical sense due to their asymmetry with respect to the different roles of the matching and matched distributions.
The Kullback-Leibler divergence is defined by 
\begin{equation}
    \label{eq:kldiv}
  S(p|q)= -\int dx \; p(x)\log \frac{p(x)}{q(x)} \;,
\end{equation}
where $p$ and $q$ are probability distributions. It is possible to symmetrize the KL divergence in order to obtain a proper metric. However in this work we have chosen to use the original KL divergence due to the different roles each of the arguments play in entropic matching: $p(x)$ is the approximative distribution matched (by changing its free parameters) to $q(x)$, which is fixed (and assumed to be correct for the purpose of matching). In this setting, $q(x)$ plays the role of a prior in phase space. Then the KL divergence represents the information loss entailed by approximating $q$ with the distribution $p$. By constraining $p$ to be Gaussian and allowing its parameters to vary freely, we obtain the Gaussian distribution with the highest information content about q. Furthermore, this functional form allows for analytical calculations and the derivation of an explicit expression for the evolution of the Gaussian parameters, whereas the symmetrized KL divergence would not.

Specifically, in our method, we match the parameters of the Gaussian $\mathcal{G}'$ to the time evolved distribution $P'$ by minimizing the KL divergence,
\begin{equation}
    \label{eq:kldivGauss}
  S(\mathcal{G}'|P')= -\int dc' \; \mathcal{G}(c'-\bar{c}',C')\log \frac{\mathcal{G}(c'-\bar{c}',C')}{P'(c')} \;.
\end{equation}
We can consider this expression as a functional to be minimized with respect to all degrees of freedom of our matching distribution, or simply as a function of the parameters $c'$ and $C'$ to be minimized with respect to their finite number of degrees of freedom. 
The minimum will define a Gaussian distribution which has maximal information content about the time evolved distribution $P'$ subject to a deterministic nonlinear evolution.

Defining $S(\mathcal{G}'|P') \equiv  S(\bar{c}',C')$, since the Kullback-Leibler divergence is now a function only of the parameters of the Gaussian, Eq.~\eqref{eq:kldiv} is explicitly 
\begin{equation}
\begin{split}
\label{eq:kldivfull}
	&S(\bar{c}',C') =\\
	& -\int dc' \; \mathcal{G}(c'-\bar{c}',C')\log \frac{\mathcal{G}(c_i'-\bar{c}',C')}{\mathcal{G}(c_i-\bar{c}(t),C(t))|_{c_i=c_i'-\delta t \dot{c}_i}\left|\frac{d c_i}{d c_i'}\right|}\\
=&\int dc' \; \mathcal{G}(c'-\bar{c}',C' ) \left[ \frac{1}{2} \log \frac{ |C'|}{|C|}+ \log\left|1-\delta t \frac{d f}{d c}\right|_{c=c'}\right.\\ 
&\left.  +\frac{1}{2} (c'-\bar{c}')^T {C'}^{-1}(c'-\bar{c}')-\frac{1}{2}(c-\bar{c})^T C^{-1}(c-\bar{c})|_{c=c'-\delta t \dot{c}}\right] \;,
\end{split}
\end{equation} 
where $P'$ was expressed as a function of $c = c' - \delta t \dot{c}$. This representation is obtained by expressing $\dot{c}$ as a finite difference, and becomes exact in the limit of $\delta t \rightarrow 0$ which is taken below to derive the differential equation. Thus the entropic divergence between the Gaussian $\mathcal{G}'$ and the time evolved former Gaussian $P'$, Eq.~\eqref{eq:kldiv}, can be brought into the form 
\begin{equation}
\begin{split}
    \label{eq:klS}
    &S(\bar{c}',C') \simeq \frac{1}{2}\log\left(\frac{|C|}{|C'|}\right)+\frac{1}{2} \text{tr}\left[ C'C^{-1} -\mathbb{1}\right.\\
    & \left.+ \delta t (2 - C^{-1} C' - C'C^{-1})\; \left\langle \frac{d f}{d c} \right\rangle_{c'} \right]\\
    &+ \frac{1}{2} (\bar{c}'-\bar{c})^T C^{-1} (\bar{c}'-\bar{c})- \delta t (\bar{c}'-\bar{c})^T C^{-1} \left\langle f(c') \right\rangle_{c'},\\
\end{split}
\end{equation}
where $\left< \:\cdot\: \right>_{c'}$ denotes the expectation value with respect to $\mathcal{G}'(c')=\mathcal{G}(c'-\bar{c}',C')$. The desired values of $\bar{c}'$ and $C'$ are obtained via minimization of Eq.~\eqref{eq:klS}, i.e. by setting the derivative of $S$ with respect to $\bar{c}'$ and $C'$ to zero, leading to  
\begin{equation}
\label{eq:params-det}
\begin{split}
    \bar{c}' &\simeq \bar{c} + \left\langle f(c)\right\rangle \delta t \\
    C' &\simeq C + \left\langle\frac{df}{dc}\right\rangle_{c}C \,\delta t + C\left\langle\frac{df}{dc}\right\rangle_{c}^T \delta t \;.
\end{split}
\end{equation}

We can now add the effect of the intrinsic noise. The noise added to the stochastic variable $c$ over the infinitesimal time $\delta t$ has a Gaussian distribution with covariance matrix $\mathcal{X}\delta t$. Therefore, the probability distribution for $c'$ including noise is the convolution of this noise Gaussian with our maximally informative Gaussian with the parameters \eqref{eq:params-det},  
\begin{align}
\label{eq:covfinal}
    P(c') &= \int d\xi \; \mathcal{G}(c'-\bar{c}' + \xi,C') \mathcal{G}(\xi,\mathcal{X}\delta t)\\
    &=\mathcal{G}(c'-\bar{c}', C'+\mathcal{X}\delta t) \;.
\end{align}
By performing this convolution and taking the limit $\delta t \rightarrow 0$, we obtain the final ODEs for the evolution of the Gaussian approximation of our system, 
\begin{equation}
\begin{split}
    \dot{\bar{c}} &= \left\langle f(c)\right\rangle \\
    \dot{C} &= \left\langle\frac{df}{dc}\right\rangle_{c}C + C\left\langle\frac{df}{dc}\right\rangle_{c}^T + \mathcal{X}  \;.
    \label{eq:solution}
\end{split}
\end{equation}
These are the general equations for our method. They are similar in form to the linear noise approximation \eqref{eq:linear} but with a key difference: In our method, the function $f(c)$ and its Jacobian are averaged over the current (Gaussian) probability distribution. As we have seen, this form follows from the Maximum Entropy principle. 

In general, the Gaussian expectation value of a nonlinear function is not analytically accessible in a closed form. One option for the numerical implementation of our method is to calculate these averages, at each time point, by numerical integration. Alternatively, one can reduce the numerical effort using analytical approximations of the averages. Towards this end, we expand the function $f(c)$ around ${\bar c}$,  
\begin{equation}
    \label{eq:taylor}
    f(c) = \sum_{n=0}^{\infty} \frac{1}{n!} \left. \frac{d^nf}{dc^n} \right|_{c={\bar c}} (c - {\bar c})^n \;,
\end{equation} 
and calculate the average of $f$ and its derivatives separately for each term,
\begin{equation}
\begin{split}
    \label{eq:taylorIndex}
    \left< f_i \right>_{c} &= f_i(\bar{c}) + \frac{1}{2} \left. \frac{d^2f_i}{dc_k \,dc_l} \right|_{c=\bar{c}} C_{kl} + \dots \\
    \left< \frac{df_i}{dc_j} \right>_{c} &= \left. \frac{df_i}{dc_j} \right|_{c=\bar{c}} + \frac{1}{2} \left. \frac{d^3f_i}{dc_j\, dc_k\, dc_l} \right|_{c=\bar{c}} C_{kl} + \dots \;,
\end{split}
\end{equation}
which is justified when the spread of the PDF around its average is not too large. If we were to take only the $n=0$ and $n=1$ terms of the expansion \eqref{eq:taylor}, we would recover the linear noise approximation \eqref{eq:linear}. However, our derivation based on the Maximum Entropy principle suggests that the linear noise approximation can be improved by keeping additional terms in this series. Our test applications below show that with only the leading corrections included, i.e. those explicitly shown in \eqref{eq:taylorIndex}, significant improvements are already obtained over the linear approximation. This is also plausible intuitively, since the leading correction introduces the feedback from the evolution of the covariance on the evolution of the mean. 

Clearly, we expect that in cases in which the dynamical equations are highly nonlinear or have several fixed points, the method will fail as the Gaussian approximation cannot accurately represent, for instance, a multimodal probability distribution. It is thus important to quantify the error made by the method. Unfortunately, as for the linear noise approximation, a simple \textit{a priori} error estimate is not readily available for our method. It appears that the only reliable way to determine whether the time evolved approximate distribution faithfully represents the exact distribution is a comparison to full stochastic simulations, which could be performed at least at a selected small set of points in the parameter space of the stochastic dynamical system. These points can be selected based upon a nonlinear dynamics analysis of the deterministic dynamical equations defined by $f$ (fixpoint and stability analysis). 

\section{Test applications}

To illustrate the benefits and limitations of our method, we now apply it to several test cases. In these applications, we integrated the ODEs of \eqref{eq:solution} with the Adams-Moulton method \cite{sewell2005numerical}. This implicit integration scheme is appropriate here due to its numerical robustness, which can successfully deal with the stiffness of the ODEs in the general nonlinear case. We compared the results to stochastic simulations of the system dynamics using the stochastic analogue of the Euler scheme, and also to the linear noise approximation \eqref{eq:linear}.

\subsection{Stochastic van der Pol oscillator}

As a convenient initial test case, we chose a system of van der Pol oscillators \cite{derPol1926}, which exhibits limit cycle behavior. Here, a single parameter controls the strength of the nonlinearity, which facilitates our study of the method's performance as a function of this nonlinearity strength. Furthermore, it has the convenient property that we can calculate the expectation values in \eqref{eq:solution} exactly, since the Taylor expansion \eqref{eq:taylor} of its function $f$ terminates at the third order. Thus, in this case our method yields the optimal Gaussian approximation (in the sense of minimal Kullback-Leibler divergence) and the only approximation consists of the fact that it enforces a Gaussian shape for the probability distribution.  

The dynamics of our stochastic van der Pol system are described by the stochastic differential equations 
\begin{equation}
\label{eq:VanDerPol}
    \ddot{x}_i = \mu(1-x_i^2)\dot{x}_i-\omega_i^2 x_i + \sum_{i\neq j} \gamma_{ij} (x_j-x_i) + \xi_i \;,
\end{equation}
where the parameter $\mu$ controls the nonlinearity, the matrix $\gamma$ controls the coupling between the different degrees of freedom (indexed by $i,j=1,\ldots,N$, with $N=3$ here), and the vector $\omega$ sets the oscillation frequency of each oscillator. We assumed constant and independent noise $\xi_{i}$ for each oscillator by taking the diagonal covariance matrix $\mathcal{X}=0.1\,\delta_{i,j}$ in all calculations and simulations. The system is transformed to a system of first order stochastic differential equations in the usual way by considering the vectors $x$ and $y= \dot{x}$ as the dynamic variables. The parameter values used in the examples were: for the coupling, $ \gamma_{i,j} = 0$ except $\gamma_{0,1} = 2$, $\gamma_{1,0} = 5$, $\gamma_{2,1} = 3$; and for the frequencies, $\omega=(2,  1,  2)$. For the nonlinearity parameter $\mu$, we used two different values, $0.05$ and $1.5$, corresponding to weak and moderate nonlinearity, respectively. As initial condition, $x(t=0) = (2,  0,  0)$ and $y(t=0)= (0,  1,  2)$ for the deterministic part and added noise according to \eqref{eq:VanDerPol}. 

\begin{figure}[htbp]
\begin{center}
\includegraphics[width=8.5cm]{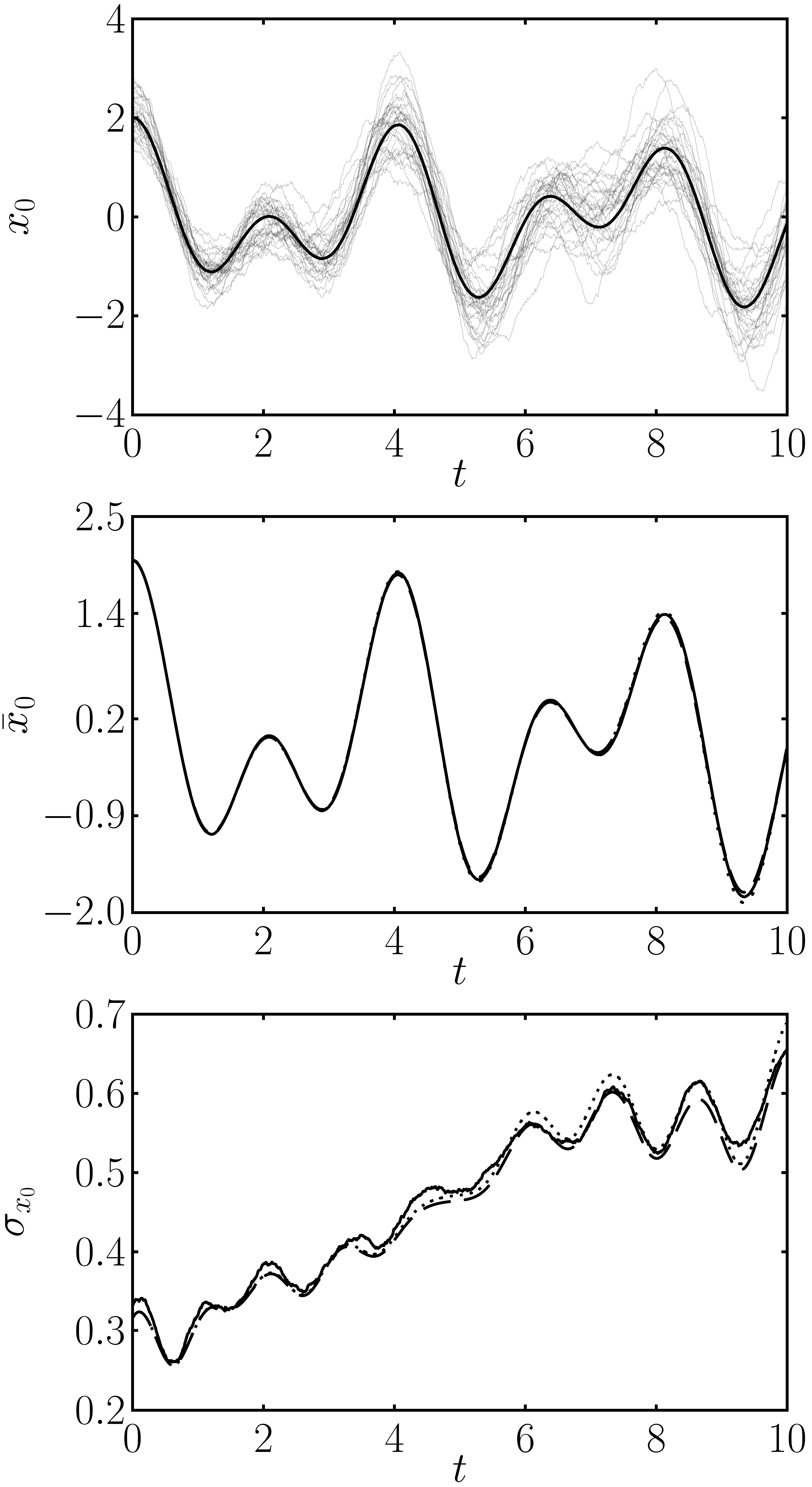}
\caption{Test application to a system of van der Pol oscillators in the regime of weak nonlinearity. Top panel: Stochastic simulation with $1000$ runs (some individual runs are plotted) of system \eqref{eq:VanDerPol} and corresponding average value for the first species of a system with $\mu=0.05$ and $N=3$. Middle panel: Comparison of stochastic average value (solid) with prediction using Eq.~\eqref{eq:solution} (dashed) and Eq.~\eqref{eq:linear} (dotted). Bottom panel: Comparison of standard deviation for the stochastic simulation (solid) with prediction using Eq.~\eqref{eq:solution} (dashed) and Eq.~\eqref{eq:linear} (dotted).}
\label{fig:osc1Traj}
\end{center}
\end{figure}

We first considered the case of weak nonlinearity ($\mu = 0.05$). 
Fig.~\ref{fig:osc1Traj} (top) shows exemplary trajectories of the coordinate $x_{0}(t)$ from our full stochastic simulations of the system \eqref{eq:VanDerPol}. The mean and variance of $x_{0}(t)$ over 1000 trajectories is shown in the middle and bottom panel, respectively, as a solid line. Superimposed are the corresponding curves for the linear noise approximation (dotted line) and our method (dashed line). With both methods, we used the same initial values for the averages and initialized the covariance matrix by the value of the intrinsic noise level. Both approximations describe the simulation very well (the agreement in the case of the mean value in the middle panel is so close that the individual curves appear indistinguishable). This behavior is expected, since for $\mu = 0.05$ the system is almost linear. 

In terms of computational effort, the full stochastic simulation consisted of $10^3$ runs of the system with time step $\delta t= 10^{-3}$. Such a small time step was necessary to ensure correct results with the Euler method. For the approximation methods, we performed a single run of the Adams-Moulton method with $\delta t = 10^{-3}$. In this case, however, decreasing $\delta t$ to $10^{-1}$ had no effect on the accuracy of the obtained trajectories. In practice, the approximation methods were two orders of magnitude faster than the stochastic simulations at the same time step (which could be safely reduced for the former, to obtain a further significant reduction of computational effort). 

\begin{figure}[htbp]
\begin{center}
\includegraphics[width=8.5cm]{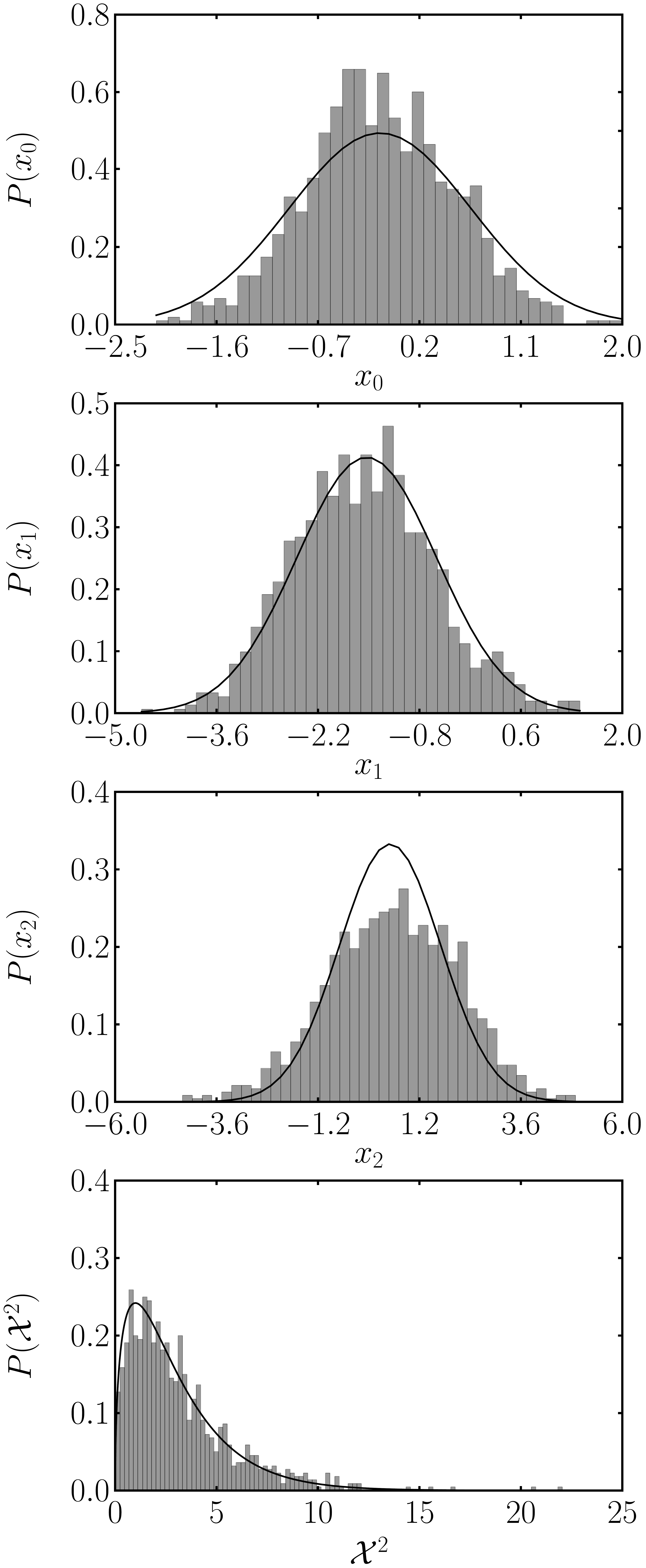}
\caption{Histogram of stochastic simulation with $1000$ runs of system \eqref{eq:VanDerPol} with $\mu=0.05$ and $N=3$ and the corresponding Gaussian (solid line) predicted by our method \eqref{eq:solution} at the final time $t=T$ (in this case $T=10$). The first three panels show the distributions of $x_{0}$, $x_{1}$, and $x_{2}$, respectively. Bottom panel: histogram of chi squared distribution for the stochastic simulation data compared to the $\chi^2$ distribution with 3 degrees of freedom.}
\label{fig:osc1Hist}
\end{center}
\end{figure}

To compare the approximations with the full simulation also on the level of the probability distributions, the first three panels of Fig.~\ref{fig:osc1Hist} show the distributions for $x_{0}$, $x_{1}$, and $x_{2}$ at time $t=10$ (histograms from the stochastic simulation. Superimposed are the Gaussian distributions (solid lines) obtained from our method, which describe the histograms well. 
Since we can visualize only projections of the full multivariate distribution, we also calculated the $\chi^2$ (chi-square) probability distribution, which can help to verify if the full distribution does indeed match a Gaussian with the predicted center and correlation matrix: If the prediction matches the simulation, a histogram of $ \chi^2 = (Y-\bar{c}_p)^\dagger C_p^{-1} (Y-\bar{c}_p)$, where $\bar{c}_p$ and $C_p$ are the predicted average and covariance, should match the chi-square distribution for a Gaussian $\mathcal{G}(\bar{c},C)$ with $N$ degrees of freedom, where $N$ is the number of dimensions of the system. The bottom panel of Fig.~\ref{fig:osc1Hist} shows that this is indeed the case for $\mu = 0.05$. 

\begin{figure}[htbp]
\begin{center}
\includegraphics[width=8.5cm]{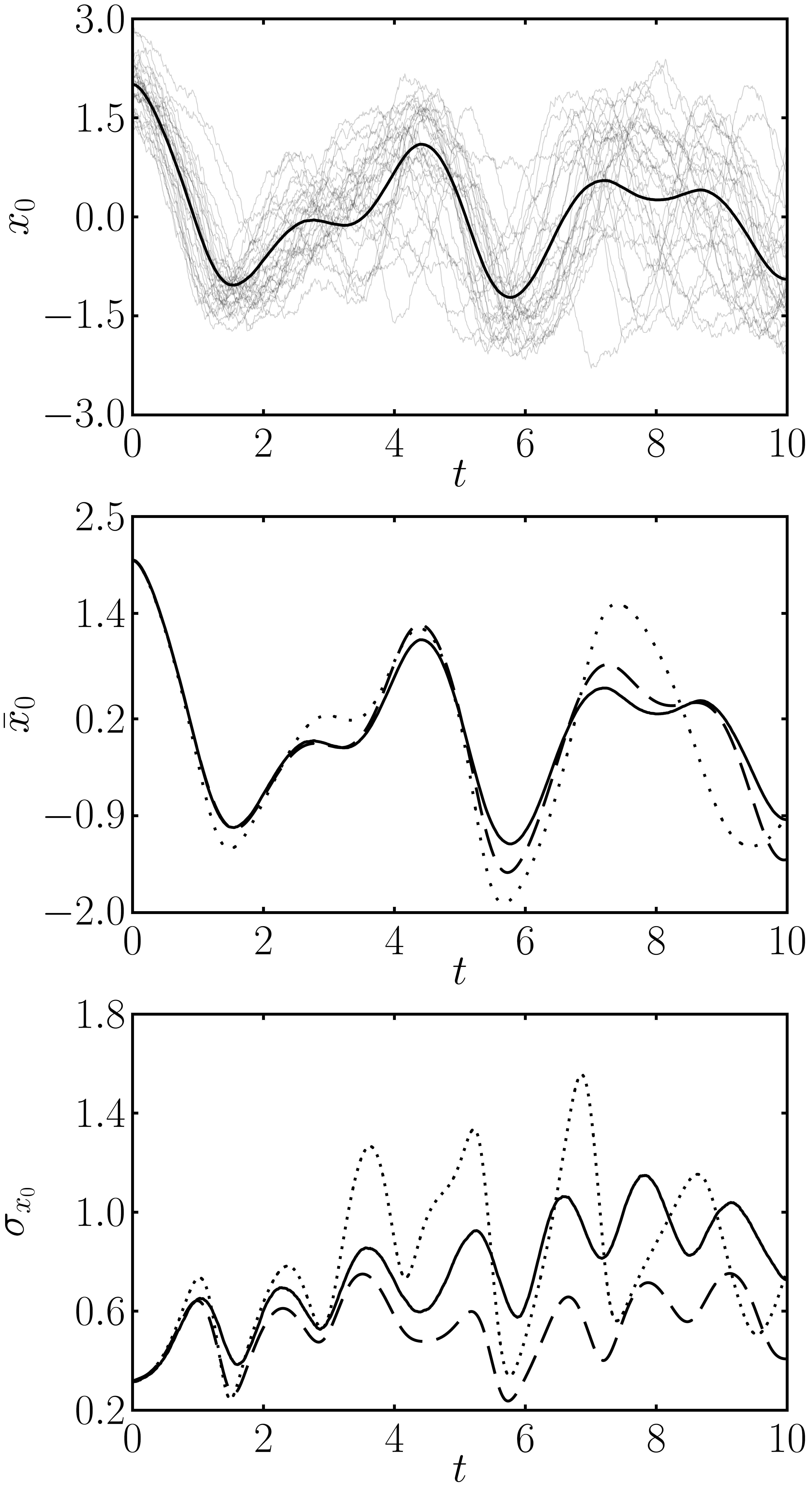}
\caption{Same as Fig.~\ref{fig:osc1Traj} but for moderately strong nonlinearity ($\mu = 1.5$).}
\label{fig:osc3Traj}
\end{center}
\end{figure}

We then considered the case of moderately strong nonlinearity ($\mu = 1.5$). Fig.~\ref{fig:osc3Traj} is the equivalent of Fig.~\ref{fig:osc1Traj} for this case. The middle panel shows that there is now a significant deviation between the mean trajectory from the simulations and the predictions of the approximation schemes. When comparing the result of the linear noise approximation (dotted line) with that of the entropically matched scheme (dashed line), it is apparent that entropic matching performs better. We attribute this to the dampening caused by progressive loss of synchronization between individual stochastic trajectories. This feedback effect from the dynamics of the covariance onto the dynamics of the mean is included only in our method. It becomes evident by working out Eq.~\eqref{eq:taylorIndex} explicitly for the van der Pol system, revealing damping terms proportional to the magnitude of the fluctuations. 

\begin{figure}[htbp]
\begin{center}
\includegraphics[width=8.5cm]{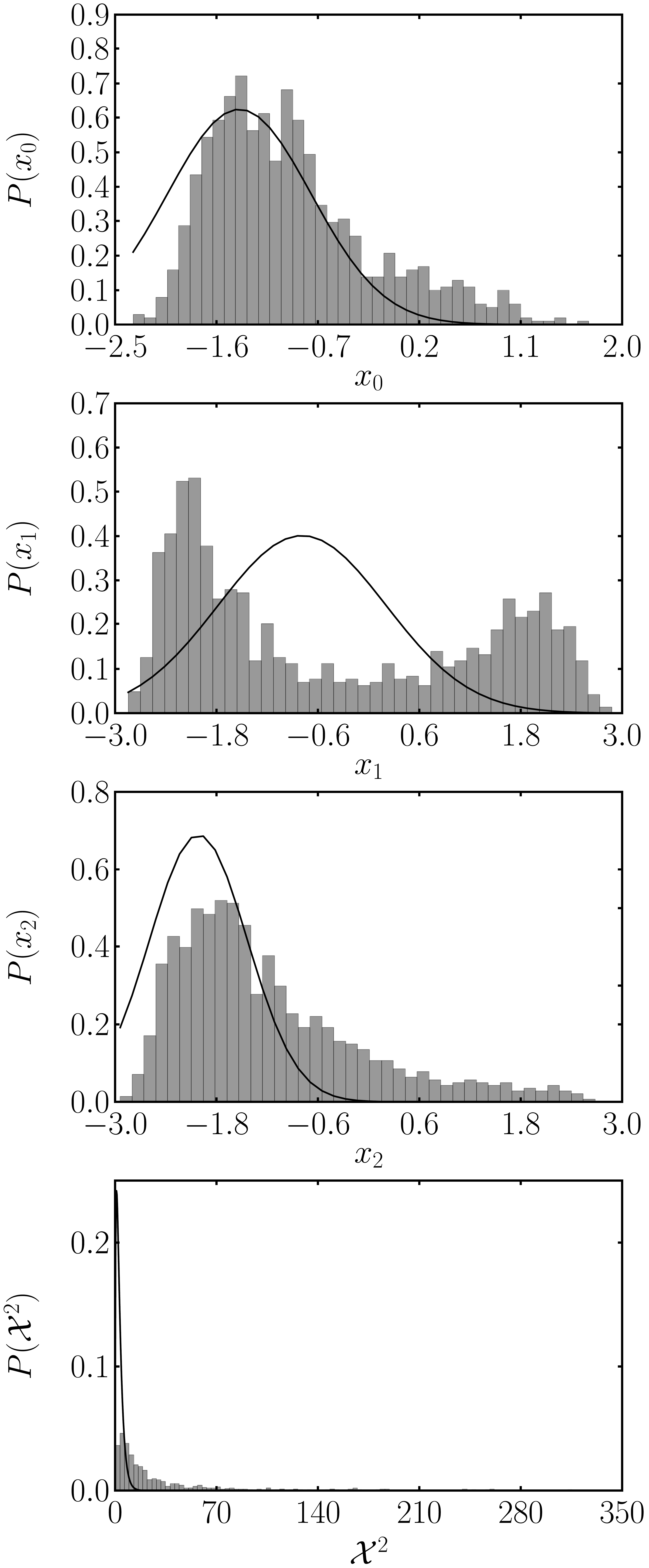}
\caption{Same as Fig.~\ref{fig:osc1Hist} but for moderately strong nonlinearity ($\mu = 1.5$).}
\label{fig:osc3Hist}
\end{center}
\end{figure}

The behavior of the variance shown in the bottom panel of Fig.~\ref{fig:osc3Traj} is consistent with this interpretation. In the linear noise approximation, the estimate for the variance strongly oscillates around the true value, under or overestimating it, until finally losing synchrony. In contrast, our method consistently underestimates the variance (inherent to the Gaussian approximation, since the true distribution exhibits more significant tails, see Fig.~\ref{fig:osc3Hist}, first panel) but always remains synchronous with the true value. Thus we consider the estimate produced by entropic matching to be more robust. Also, we found that in some parameter regimes, the covariance matrix diverges within the linear noise approximation while the entropically matched scheme still produces finite estimates. 

Fig.~\ref{fig:osc3Hist} shows the distributions as in Fig.~\ref{fig:osc1Hist}, now for our case of moderately strong nonlinearity. We note that the distribution remains unimodal for two components of the system while it becomes bimodal for one. Even though the unimodal distributions are non Gaussian our method still approximates their average and standard deviation well, which is the intended behavior. In the chi-square distribution of the bottom panel, it is also visible that the multivariate Gaussian no longer captures the shape of the distribution. This illustrates that the chi-square histogram is indeed a good indicator for the extent to which a multidimensional distribution has Gaussian shape. More generally, the chi-square distribution can be a very convenient measure for the quality at which our approximation describes the full stochastic dynamics of a higher-dimensional system. 

Taken together, this test application has illustrated that our method can still lead to a fair description of a stochastic dynamical system with moderately strong nonlinearity, considerably better than the linear noise approximation at similar computational effort, and much less computational effort than full simulation.

\subsection{Genetic circuit model} 

As a second test application, we studied a simple model for a genetic circuit. The model consists of a system of mutually repressing genes inspired by the ``repressilator'' \cite{elowitz_synthetic_2000}, which exhibits noisy oscillations in a certain range of the parameter regime. In the model, 
gene regulation is described by Hill-type regulation functions, where each gene is repressed by another in the circuit, such that 
\begin{equation}
\label{eq:geneNet}
\dot{c}_i = \frac{k^n}{c_j^n+k^n} - \lambda c_i + \zeta_i \;,
\end{equation}
where $j = (i+1)\mod N$ (and $\mod$ represents the modulo operator), $k$ is the expression level of the input gene at which the target is repressed by $50\,\%$, and $n$ is the binding cooperativity or Hill exponent (with increasing $n$ making the regulation more nonlinear, i.e. step-like; in the following, we assume $n=2$). 

For the case of gene regulation, the noise cannot be assumed to be constant, but rather scales with the mean level. In the case of intrinsic noise, which we assume here, $\sigma^2_c\propto \langle c\rangle$, i.e. the variance increases linearly with the mean \cite{swain_intrinsic_2002}. As mentioned above, we assume concentrations large enough such that $\zeta_i$ can be regarded as continuous. In order to enable us to apply the exact same mathematical framework as in the previous section, we perform a change of variables by introducing a new variable $\rho = \log c$ such that
\begin{equation}
\label{eq:loggeneNet}
\dot{\rho}_i = \frac{k^n e^{-\rho_i}}{e^{\rho_j n}+k^n} - \lambda + \xi_i
\end{equation}
where now the noise $\xi$ has constant variance ($0.1$ in our numerical example). We again assume a Gaussian white noise process for $\xi$, which means that the distribution of $\zeta_i$ obtains a log-normal shape. Note that the precise shape of the distribution is not important here, since we are not trying to describe a specific experiment, but use \eqref{eq:geneNet} as a toy model for illustration. 

\begin{figure}[htbp]
\begin{center}
\includegraphics[width=0.5\textwidth]{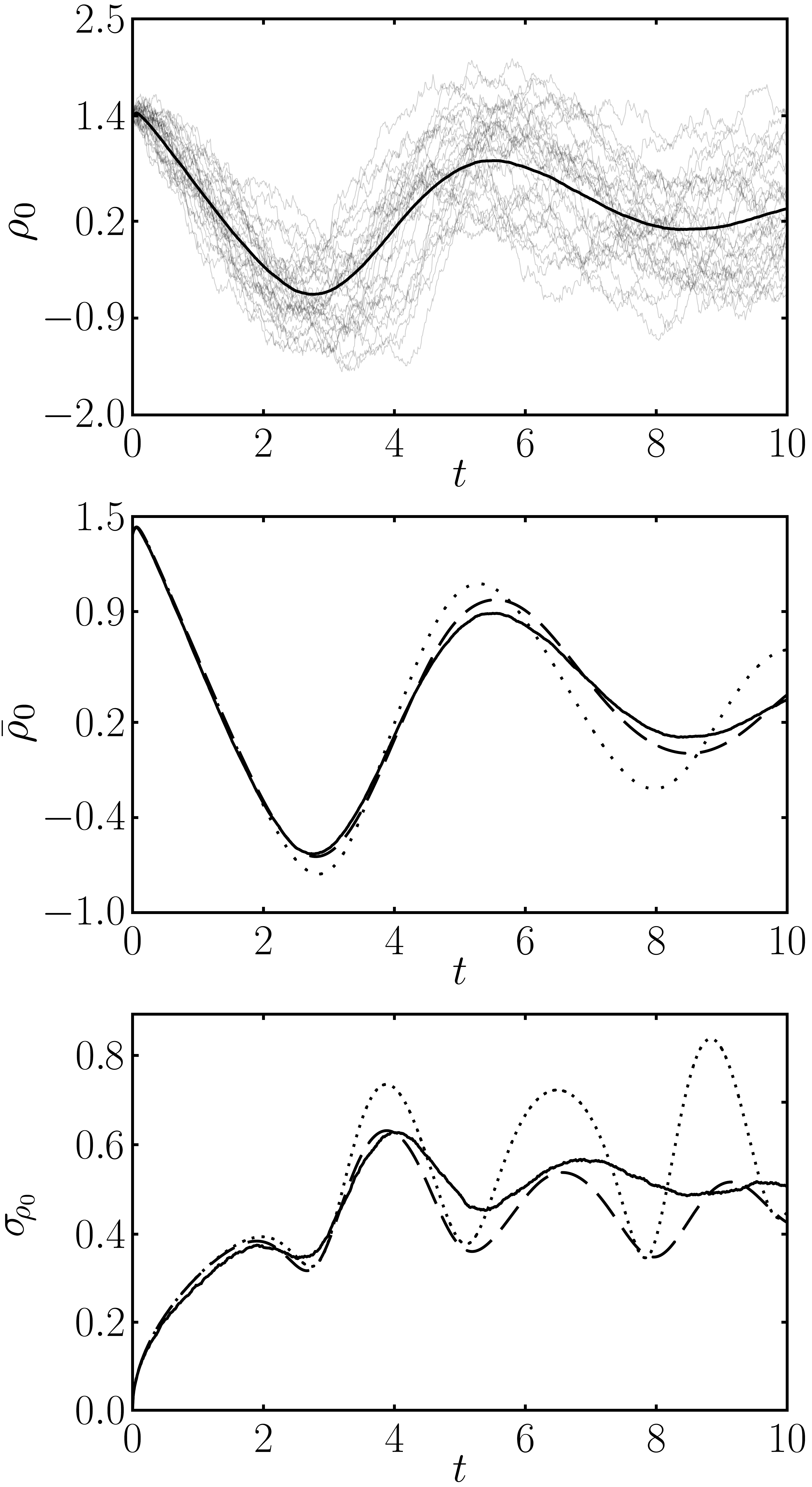}
\caption{Analogous to Fig.~\ref{fig:osc1Hist} but for the genetic circuit model \eqref{eq:loggeneNet} with $n=2$ and $N=3$.}
\label{fig:geneTraj}
\end{center}
\end{figure}

In contrast to the previous test application, our approximation \eqref{eq:taylorIndex} of the averages (again truncated after the leading correction to the linear noise approximation) is not exact in this example. Therefore, the genetic circuit example also serves us to test whether this approximation significantly degrades the benefits of the method. Fig.~\ref{fig:geneTraj} characterizes the stochastic nonlinear dynamics of this model and the performance of the approximation schemes in the same way as Fig.~\ref{fig:osc1Traj} does for the van der Pol oscillator system. The exemplary individual trajectories from the stochastic simulation, shown in the top panel, clearly convey how significant the noise is in the regime of our simulation. 

\begin{figure}[htbp]
\begin{center}
\includegraphics[width=0.5\textwidth]{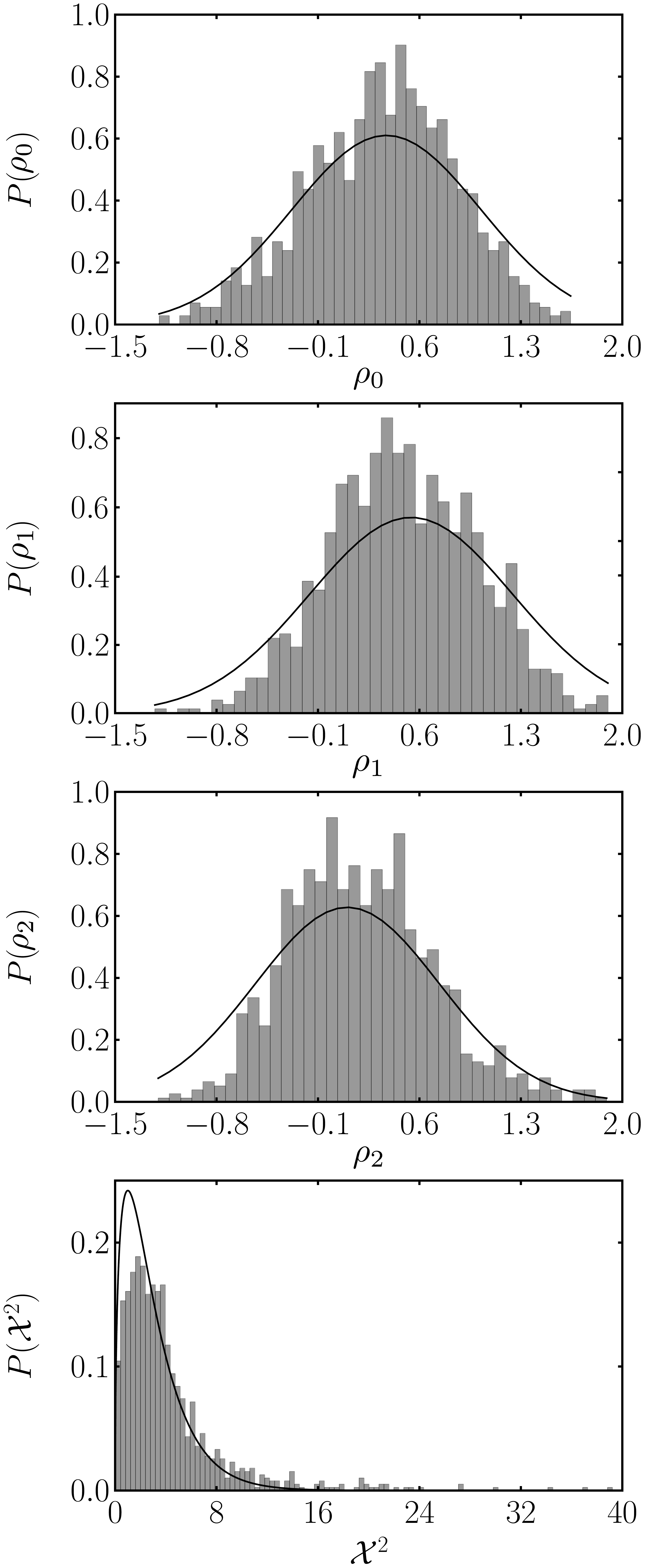}
\caption{Final histogram of stochastic simulation with $1000$ runs of 
genetic circuit model \eqref{eq:loggeneNet} with $n=2$ and $N=3$. }
\label{fig:geneHist}
\end{center}
\end{figure}

The middle and bottom panels show that entropic matching again leads to a better description of the simulation than the linear noise approximation, with similar behaviour as in the previous case. This indicates that enforcing the probability distribution to be Gaussian constitutes a more drastic approximation than neglecting the higher sub-leading orders in the expansion \eqref{eq:taylorIndex}. From the histograms in Fig.~\ref{fig:geneHist} we can see that the distributions remain unimodal, but are slightly asymmetrical. However, the Gaussian approximation is still relatively good, as evidenced also by the chi-squared distribution in the bottom panel. 

\section{Discussion and Conclusions}

This paper proposes an approximation method to predict the evolution of stochastic nonlinear systems such as biochemical networks. The method is based on entropic matching strategy \cite{2012arXiv1206.4229E} developed to construct simulation schemes based on the information theoretical concept of the Maximum Entropy Principle. For each step in the time evolution of a system, the probability distribution describing the state of the system is matched to a Gaussian and ODEs for the first and second moments of the distribution are derived (Eq.~\ref{eq:solution}). The evolution of these moments then tracks the average and covariance of the system parameters. We have implemented these equations numerically employing a series expansion to compute the Gaussian expectation values in \eqref{eq:solution}. We found that this scheme leads to a more accurate approximation method, which is useful in a wider parameter regime than the linear noise approximation. In the cases where the system's intrinsic noise is very low or the network dynamics are approximately linear, both methods work equally well. For moderate noise and weak nonlinearity, our method outperforms the linear noise approximation. 

Just as the linear noise approximation, our method is much faster than full stochastic simulation due to a number of factors. To begin with, numerical algorithms for the solution of SDEs cannot guarantee convergence as fast as standard ODE routines. As an example, the commonly used stochastic Euler scheme has an order of convergence of merely $0.5$ whereas the Adams-Moulton method we use has an order of convergence of $4$. Algorithms with even faster convergence orders are available and simple to implement. On the other hand, more advanced algorithms for SDEs yield marginally better convergence at the cost of high complexity in terms of implementation~\cite{kloeden_numerical_1992}. The upshot is that much smaller time steps must be used for a stochastic simulation, thereby increasing run time. A second important point is the issue of sampling statistics. To sample from the probability distribution of interest with good accuracy, one must obtain thousands of paths from the SDE simulation with variance scaling as $1/\sqrt{N_{\text{runs}}}$. Importance sampling can mitigate this problem somewhat, but the number of paths required is still very high. On the other hand, the linear noise approximation and our scheme both require only one evaluation.

A complication with our method (compared to the linear noise approximation) is the calculation of higher order derivatives. However, algorithmic differentiation methods\cite{griewank_evaluating_2008} allow to calculate derivative tensors of any order very efficiently. The calculation of these tensors is comparable in complexity to the calculation of the original function, being slower by only a fixed constant depending on the order of the derivative.

In the case of strong nonlinearities, stochastic simulations are still the preferred method. Clearly, multimodal or highly asymmetric target distributions will be poorly approximated by a multivariate Gaussian. To remedy this, a possibility would be the use of a Gaussian mixture to represent the different modes of the probability distribution. However, this approach brings with it a number of problems as the entropic divergence cannot be computed analytically in this case. Additionally the number of parameters to be determined would scale at least as $L \, (N + \frac{N(N+1)}{2} + 1)$, where $L$ is the number of Gaussian mixture components. This can become unfeasible for a large number of mixture components.

We envisage that an important application of our method will be the reverse engineering of genetic regulation networks (see, e.g. \cite{diambra_coarse-grain_2011}), where the values for the parameters $a$ will represent the network structure and kinetic parameters. To infer the parameter values from the data, it is necessary to calculate the posterior density  
\begin{equation}
\label{eq:bayesbase}
P( a| c, t) =  \frac{P(a) \prod_{j}^{J} P(c | a, t_j)}{\int da \; P(a) \prod_{j}^{J} P(c | a, t_j)}
\end{equation}
where $P(a)$ represents the prior knowledge about the parameters of the system. 
Thus, the posterior density can be obtained once we have calculated the likelihood or ``forward probability'' $P(c|a,t_j)$ for all relevant time points. The proposed method may be sufficient to derive an efficient inference scheme for stochastic nonlinear networks, an issue left for the future.


\begin{thebibliography}{30}
\expandafter\ifx\csname natexlab\endcsname\relax\def\natexlab#1{#1}\fi
\expandafter\ifx\csname bibnamefont\endcsname\relax
  \def\bibnamefont#1{#1}\fi
\expandafter\ifx\csname bibfnamefont\endcsname\relax
  \def\bibfnamefont#1{#1}\fi
\expandafter\ifx\csname citenamefont\endcsname\relax
  \def\citenamefont#1{#1}\fi
\expandafter\ifx\csname url\endcsname\relax
  \def\url#1{\texttt{#1}}\fi
\expandafter\ifx\csname urlprefix\endcsname\relax\def\urlprefix{URL }\fi
\providecommand{\bibinfo}[2]{#2}
\providecommand{\eprint}[2][]{\url{#2}}

\bibitem[{\citenamefont{Raser and O'Shea}(2005)}]{Raser:2005gr}
\bibinfo{author}{\bibfnamefont{J.~M.} \bibnamefont{Raser}} \bibnamefont{and}
  \bibinfo{author}{\bibfnamefont{E.~K.} \bibnamefont{O'Shea}},
  \bibinfo{journal}{Science} \textbf{\bibinfo{volume}{309}},
  \bibinfo{pages}{2010} (\bibinfo{year}{2005}).

\bibitem[{\citenamefont{Eldar and Elowitz}(2010)}]{Eldar:2010kk}
\bibinfo{author}{\bibfnamefont{A.}~\bibnamefont{Eldar}} \bibnamefont{and}
  \bibinfo{author}{\bibfnamefont{M.~B.} \bibnamefont{Elowitz}},
  \bibinfo{journal}{Nature} \textbf{\bibinfo{volume}{467}},
  \bibinfo{pages}{167} (\bibinfo{year}{2010}).

\bibitem[{\citenamefont{Bal{\'a}zsi et~al.}(2011)\citenamefont{Bal{\'a}zsi, van
  Oudenaarden, and Collins}}]{Balazsi:2011bw}
\bibinfo{author}{\bibfnamefont{G.}~\bibnamefont{Bal{\'a}zsi}},
  \bibinfo{author}{\bibfnamefont{A.}~\bibnamefont{van Oudenaarden}},
  \bibnamefont{and} \bibinfo{author}{\bibfnamefont{J.~J.}
  \bibnamefont{Collins}}, \bibinfo{journal}{Cell}
  \textbf{\bibinfo{volume}{144}}, \bibinfo{pages}{910} (\bibinfo{year}{2011}).

\bibitem[{\citenamefont{Munsky et~al.}(2012)\citenamefont{Munsky, Neuert, and
  van Oudenaarden}}]{Munsky:2012ie}
\bibinfo{author}{\bibfnamefont{B.}~\bibnamefont{Munsky}},
  \bibinfo{author}{\bibfnamefont{G.}~\bibnamefont{Neuert}}, \bibnamefont{and}
  \bibinfo{author}{\bibfnamefont{A.}~\bibnamefont{van Oudenaarden}},
  \bibinfo{journal}{Science} \textbf{\bibinfo{volume}{336}},
  \bibinfo{pages}{183} (\bibinfo{year}{2012}).

\bibitem[{\citenamefont{Elowitz et~al.}(2002)\citenamefont{Elowitz, Levine,
  Siggia, and Swain}}]{Elowitz:2002wq}
\bibinfo{author}{\bibfnamefont{M.}~\bibnamefont{Elowitz}},
  \bibinfo{author}{\bibfnamefont{A.}~\bibnamefont{Levine}},
  \bibinfo{author}{\bibfnamefont{E.}~\bibnamefont{Siggia}}, \bibnamefont{and}
  \bibinfo{author}{\bibfnamefont{P.}~\bibnamefont{Swain}},
  \bibinfo{journal}{Science} \textbf{\bibinfo{volume}{297}},
  \bibinfo{pages}{1183} (\bibinfo{year}{2002}).

\bibitem[{\citenamefont{de~Jong}(2002)}]{de_jong_modeling_2002}
\bibinfo{author}{\bibfnamefont{H.}~\bibnamefont{de~Jong}}, \bibinfo{journal}{J.
  Comp. Biol.} \textbf{\bibinfo{volume}{9}}, \bibinfo{pages}{67}
  (\bibinfo{year}{2002}).

\bibitem[{\citenamefont{Wilkinson}(2009)}]{wilkinson_stochastic_2009}
\bibinfo{author}{\bibfnamefont{D.~J.} \bibnamefont{Wilkinson}},
  \bibinfo{journal}{Nat. Rev. Genet.} \textbf{\bibinfo{volume}{10}},
  \bibinfo{pages}{122} (\bibinfo{year}{2009}).

\bibitem[{\citenamefont{{van Kampen}}(2007)}]{van_kampen_2007}
\bibinfo{author}{\bibfnamefont{N.~G.} \bibnamefont{{van Kampen}}},
  \emph{\bibinfo{title}{Stochastic Processes In Physics And Chemistry}}
  (\bibinfo{publisher}{Elsevier Science and Technology Books},
  \bibinfo{year}{2007}), \bibinfo{edition}{3rd} ed.

\bibitem[{\citenamefont{Gardiner}(2010)}]{gardiner_stochastic_2010}
\bibinfo{author}{\bibfnamefont{C.}~\bibnamefont{Gardiner}},
  \emph{\bibinfo{title}{Stochastic Methods: A Handbook for the Natural and
  Social Sciences}} (\bibinfo{publisher}{Springer}, \bibinfo{year}{2010}).

\bibitem[{\citenamefont{S{\"u}el et~al.}(2006)\citenamefont{S{\"u}el,
  Garcia-Ojalvo, Liberman, and Elowitz}}]{Suel:2006ea}
\bibinfo{author}{\bibfnamefont{G.~M.} \bibnamefont{S{\"u}el}},
  \bibinfo{author}{\bibfnamefont{J.}~\bibnamefont{Garcia-Ojalvo}},
  \bibinfo{author}{\bibfnamefont{L.~M.} \bibnamefont{Liberman}},
  \bibnamefont{and} \bibinfo{author}{\bibfnamefont{M.~B.}
  \bibnamefont{Elowitz}}, \bibinfo{journal}{Nature}
  \textbf{\bibinfo{volume}{440}}, \bibinfo{pages}{545} (\bibinfo{year}{2006}).

\bibitem[{\citenamefont{Elowitz and Leibler}(2000)}]{elowitz_synthetic_2000}
\bibinfo{author}{\bibfnamefont{M.~B.} \bibnamefont{Elowitz}} \bibnamefont{and}
  \bibinfo{author}{\bibfnamefont{S.}~\bibnamefont{Leibler}},
  \bibinfo{journal}{Nature} \textbf{\bibinfo{volume}{403}},
  \bibinfo{pages}{335} (\bibinfo{year}{2000}).

\bibitem[{\citenamefont{Garcia-Ojalvo et~al.}(2004)\citenamefont{Garcia-Ojalvo,
  Elowitz, and Strogatz}}]{GarciaOjalvo:2004vg}
\bibinfo{author}{\bibfnamefont{J.}~\bibnamefont{Garcia-Ojalvo}},
  \bibinfo{author}{\bibfnamefont{M.~B.} \bibnamefont{Elowitz}},
  \bibnamefont{and} \bibinfo{author}{\bibfnamefont{S.~H.}
  \bibnamefont{Strogatz}}, \bibinfo{journal}{Proc. Natl. Acad. Sci.}
  \textbf{\bibinfo{volume}{101}}, \bibinfo{pages}{10955}
  (\bibinfo{year}{2004}).

\bibitem[{\citenamefont{Prill et~al.}(2011)\citenamefont{Prill, Saez-Rodriguez,
  Alexopoulos, Sorger, and Stolovitzky}}]{Prill:2011cc}
\bibinfo{author}{\bibfnamefont{R.~J.} \bibnamefont{Prill}},
  \bibinfo{author}{\bibfnamefont{J.}~\bibnamefont{Saez-Rodriguez}},
  \bibinfo{author}{\bibfnamefont{L.~G.} \bibnamefont{Alexopoulos}},
  \bibinfo{author}{\bibfnamefont{P.~K.} \bibnamefont{Sorger}},
  \bibnamefont{and}
  \bibinfo{author}{\bibfnamefont{G.}~\bibnamefont{Stolovitzky}},
  \bibinfo{journal}{Sci. Signal.} \textbf{\bibinfo{volume}{4}},
  \bibinfo{pages}{mr7} (\bibinfo{year}{2011}).

\bibitem[{\citenamefont{Steiert et~al.}(2012)\citenamefont{Steiert, Raue,
  Timmer, and Kreutz}}]{Steiert:2012ke}
\bibinfo{author}{\bibfnamefont{B.}~\bibnamefont{Steiert}},
  \bibinfo{author}{\bibfnamefont{A.}~\bibnamefont{Raue}},
  \bibinfo{author}{\bibfnamefont{J.}~\bibnamefont{Timmer}}, \bibnamefont{and}
  \bibinfo{author}{\bibfnamefont{C.}~\bibnamefont{Kreutz}},
  \bibinfo{journal}{PLoS ONE} \textbf{\bibinfo{volume}{7}},
  \bibinfo{pages}{e40052} (\bibinfo{year}{2012}).

\bibitem[{\citenamefont{Walczak et~al.}(2009)\citenamefont{Walczak, Mugler, and
  Wiggins}}]{walczak_stochastic_2009}
\bibinfo{author}{\bibfnamefont{A.~M.} \bibnamefont{Walczak}},
  \bibinfo{author}{\bibfnamefont{A.}~\bibnamefont{Mugler}}, \bibnamefont{and}
  \bibinfo{author}{\bibfnamefont{C.~H.} \bibnamefont{Wiggins}},
  \bibinfo{journal}{Proc. Natl. Acad. Sci.} \textbf{\bibinfo{volume}{106}},
  \bibinfo{pages}{6529} (\bibinfo{year}{2009}).

\bibitem[{\citenamefont{Barzel and Biham}(2011)}]{barzel_binomial_2011}
\bibinfo{author}{\bibfnamefont{B.}~\bibnamefont{Barzel}} \bibnamefont{and}
  \bibinfo{author}{\bibfnamefont{O.}~\bibnamefont{Biham}},
  \bibinfo{journal}{Phys. Rev. Lett.} \textbf{\bibinfo{volume}{106}},
  \bibinfo{pages}{150602} (\bibinfo{year}{2011}).

\bibitem[{\citenamefont{Allen et~al.}(2005)\citenamefont{Allen, Warren, and
  Ten~Wolde}}]{allen_sampling_2005}
\bibinfo{author}{\bibfnamefont{R.~J.} \bibnamefont{Allen}},
  \bibinfo{author}{\bibfnamefont{P.~B.} \bibnamefont{Warren}},
  \bibnamefont{and} \bibinfo{author}{\bibfnamefont{P.~R.}
  \bibnamefont{Ten~Wolde}}, \bibinfo{journal}{Phys. Rev. Lett.}
  \textbf{\bibinfo{volume}{94}}, \bibinfo{pages}{18104} (\bibinfo{year}{2005}).

\bibitem[{\citenamefont{Becker et~al.}(2012)\citenamefont{Becker, Allen, and
  ten Wolde}}]{Becker:2012fl}
\bibinfo{author}{\bibfnamefont{N.~B.} \bibnamefont{Becker}},
  \bibinfo{author}{\bibfnamefont{R.~J.} \bibnamefont{Allen}}, \bibnamefont{and}
  \bibinfo{author}{\bibfnamefont{P.~R.} \bibnamefont{ten Wolde}},
  \bibinfo{journal}{J. Chem. Phys.} \textbf{\bibinfo{volume}{136}},
  \bibinfo{pages}{174118} (\bibinfo{year}{2012}).

\bibitem[{\citenamefont{{Kullback} and {Leibler}}(1951)}]{Kullback1951}
\bibinfo{author}{\bibfnamefont{S.}~\bibnamefont{{Kullback}}} \bibnamefont{and}
  \bibinfo{author}{\bibfnamefont{R.}~\bibnamefont{{Leibler}}},
  \bibinfo{journal}{Ann. Math. Statist.} \textbf{\bibinfo{volume}{22}},
  \bibinfo{pages}{79} (\bibinfo{year}{1951}).

\bibitem[{\citenamefont{{En{\ss}lin}}(2012)}]{2012arXiv1206.4229E}
\bibinfo{author}{\bibfnamefont{T.~A.} \bibnamefont{{En{\ss}lin}}},
  \bibinfo{journal}{accepted by Phys. Rev. E}  (\bibinfo{year}{2012}),
  \eprint{arXiv:1206.4229}.

\bibitem[{\citenamefont{Kloeden and Platen}(1992)}]{kloeden_numerical_1992}
\bibinfo{author}{\bibfnamefont{P.~E.} \bibnamefont{Kloeden}} \bibnamefont{and}
  \bibinfo{author}{\bibfnamefont{E.}~\bibnamefont{Platen}},
  \emph{\bibinfo{title}{Numerical Solution of Stochastic Differential
  Equations}} (\bibinfo{publisher}{Springer}, \bibinfo{year}{1992}),
  \bibinfo{edition}{corrected} ed.

\bibitem[{\citenamefont{Jaynes}(1957{\natexlab{a}})}]{1957PhRv..106..620J}
\bibinfo{author}{\bibfnamefont{E.~T.} \bibnamefont{Jaynes}},
  \bibinfo{journal}{Phys. Rev.} \textbf{\bibinfo{volume}{106}},
  \bibinfo{pages}{620} (\bibinfo{year}{1957}{\natexlab{a}}).

\bibitem[{\citenamefont{Jaynes}(1957{\natexlab{b}})}]{1957PhRv..108..171J}
\bibinfo{author}{\bibfnamefont{E.~T.} \bibnamefont{Jaynes}},
  \bibinfo{journal}{Phys. Rev.} \textbf{\bibinfo{volume}{108}},
  \bibinfo{pages}{171} (\bibinfo{year}{1957}{\natexlab{b}}).

\bibitem[{\citenamefont{{Jaynes} and {Bretthorst}}(2003)}]{2003prth.book.....J}
\bibinfo{author}{\bibfnamefont{E.~T.} \bibnamefont{{Jaynes}}} \bibnamefont{and}
  \bibinfo{author}{\bibfnamefont{G.~L.} \bibnamefont{{Bretthorst}}},
  \emph{\bibinfo{title}{{Probability Theory}}} (\bibinfo{publisher}{Cambridge
  University Press}, \bibinfo{year}{2003}).

\bibitem[{\citenamefont{{En{\ss}lin} and {Weig}}(2010)}]{2010PhRvE..82e1112E}
\bibinfo{author}{\bibfnamefont{T.~A.} \bibnamefont{{En{\ss}lin}}}
  \bibnamefont{and} \bibinfo{author}{\bibfnamefont{C.}~\bibnamefont{{Weig}}},
  \bibinfo{journal}{\pre} \textbf{\bibinfo{volume}{82}}, \bibinfo{eid}{051112}
  (\bibinfo{year}{2010}).

\bibitem[{\citenamefont{Sewell}(2005)}]{sewell2005numerical}
\bibinfo{author}{\bibfnamefont{G.}~\bibnamefont{Sewell}},
  \emph{\bibinfo{title}{The Numerical Solution of Ordinary and Partial
  Differential Equations}}, Pure and Applied Mathematics Series
  (\bibinfo{publisher}{Wiley}, \bibinfo{year}{2005}).

\bibitem[{\citenamefont{van~der Pol}(1926)}]{derPol1926}
\bibinfo{author}{\bibfnamefont{B.}~\bibnamefont{van~der Pol}},
  \bibinfo{journal}{Philosophical Magazine Series 7}
  \textbf{\bibinfo{volume}{2}}, \bibinfo{pages}{978} (\bibinfo{year}{1926}).

\bibitem[{\citenamefont{Swain et~al.}(2002)\citenamefont{Swain, Elowitz, and
  Siggia}}]{swain_intrinsic_2002}
\bibinfo{author}{\bibfnamefont{P.~S.} \bibnamefont{Swain}},
  \bibinfo{author}{\bibfnamefont{M.~B.} \bibnamefont{Elowitz}},
  \bibnamefont{and} \bibinfo{author}{\bibfnamefont{E.~D.}
  \bibnamefont{Siggia}}, \bibinfo{journal}{Proc. Natl. Acad. Sci.}
  \textbf{\bibinfo{volume}{99}}, \bibinfo{pages}{12795} (\bibinfo{year}{2002}).

\bibitem[{\citenamefont{Griewank and Walther}(2008)}]{griewank_evaluating_2008}
\bibinfo{author}{\bibfnamefont{A.}~\bibnamefont{Griewank}} \bibnamefont{and}
  \bibinfo{author}{\bibfnamefont{A.}~\bibnamefont{Walther}},
  \emph{\bibinfo{title}{Evaluating Derivatives: Principles and Techniques of
  Algorithmic Differentiation}} (\bibinfo{publisher}{{SIAM}},
  \bibinfo{year}{2008}).

\bibitem[{\citenamefont{Diambra}(2011)}]{diambra_coarse-grain_2011}
\bibinfo{author}{\bibfnamefont{L.}~\bibnamefont{Diambra}},
  \bibinfo{journal}{Phys. A} \textbf{\bibinfo{volume}{390}},
  \bibinfo{pages}{2198} (\bibinfo{year}{2011}).

\end{thebibliography}
\end{document}